\documentstyle[aps,prb,epsf]{revtex}
\begin{document}
\draft
\title{Cumulant approach to the low-temperature thermodynamics of 
       many-body systems} 
\author{Holger K\"{o}hler, Matthias Vojta, and Klaus W.~Becker}
\address{Institut f\"{u}r Theoretische Physik,
Technische Universit\"{a}t Dresden, D-01062 Dresden, \\ Germany}
\maketitle

\begin{abstract} 
Current methods to describe the thermodynamic behavior
of many-particle systems are often based on  
perturbation theory with an unperturbed system consisting of
free particles. Therefore, only a few methods are
able to describe both strongly and weakly correlated systems
along the same lines. In this article we propose a cumulant approach
which allows for the evaluation of excitation energies and is
especially appropriate to account for the thermodynamics at 
low temperatures. The method is an extension of
a cumulant formalism which was recently proposed to study
statical and dynamical properties of many-body systems at zero
temperature. The present approach merges into the former
one for vanishing temperature. 
As an application we investigate the thermodynamics of the hole-doped 
antiferromagnetic phase in high-temperature superconductors in the 
framework of the anisotropic $t$-$J$ model.\\
\end{abstract}



\section{Introduction} 
For the investigation of many-particle systems at finite temperatures 
the free energy $F$ plays a central role. One basic property of the free 
energy is its size consistency, i.e. the free energy scales with the 
size of the system. Each approximation which is used to evaluate $F$ 
must preserve this property. In diagrammatic approaches size 
consistency is guaranteed by the fact that in any physical quantity  
only linked diagrams enter. Usually, diagram technique makes use of
Wick's theorem, which is appropriate only if the unperturbed Hamiltonian 
is a single-particle Hamiltonian. Therefore standard diagrammatic 
approaches are restricted to weakly correlated systems. 
An alternative approach 
to evaluate statical and dynamical properties at zero 
temperature was recently proposed \cite{BeckFul88,BeckWonFul89,BeckBre90} and 
is based on the introduction of cumulants.
As is well known from statistical physics the use of cumulants ensures
size consistency. This favors the cumulant method 
especially for the description of strongly correlated 
systems though it may be applied as well to weakly correlated
systems. The main aspect of the present paper is to investigate excitation 
energies which are needed to evaluate 
the free energy. 

There is a natural danger that a cumulant expansion of 
$F$ is only valid for high temperatures. Expanding 
with respect to small values of the inverse temperature 
$\beta= 1/kT$ the partition function reads
\begin{eqnarray}
\label{1}
   Z=Z\left(0\right) \langle 1 -\beta H +\frac{1}{2}\beta^{2} H^{2}
     \cdots \rangle_{0}\,\,\,,
\end{eqnarray}
where $\langle\cdots\rangle_{0}=Tr(\dots)/Tr\,1$
stands for an unweighted average at infinite temperature 
and $Z\left(0\right)=Tr\,1$ is the trace of the unity operator $1$.
Expanding $\ln{Z}$, suitably 
regrouped into contributions in powers of $\beta$, one finds
for the free energy
\begin{eqnarray}
\label{2}  
 F=&& -\frac{1}{\beta}\ln Z\left(0\right) +\langle H \rangle_{0} +
	 \frac{1}{2}\beta \left[ \langle H^2 \rangle_{0}-
				 \langle H \rangle^{2}_{0} \right] 
				 \\
        &&+ \frac{\beta^2}{3!}\left[\langle H^{3} \rangle_{0}-
			 3 \langle H^{2} \rangle_{0} \langle H \rangle_{0}+
			 2 \langle H^{3} \rangle_{0} \right] 
			 + \cdots\,\,\,, \nonumber
\end{eqnarray}
Note that the expressions in the brackets $\left[\cdots\right]$ are 
cumulants.

In the present approach we want to avoid that a 
free energy expansion like (\ref{2}) is only valid for high 
temperatures. Of more interest
are low-temperature expansions. A low-temperature expansion for
$F$ must reduce to the ground-state energy in the limit $ T\rightarrow 0$. 
In contrast to a high-temperature expansion where $\beta$ is small, for 
sufficiently low temperatures the canonical weight 
$\exp\left(-\beta (E_{n}-E_0)\right)$ 
is small for excitations $E_{n}$ higher than the ground-state 
energy $E_0$. Therefore, in case of a non-degenerate ground state 
the free energy can be expanded for low temperatures as follows
\begin{eqnarray}
\label{3}
  F&=& -\frac{1}{\beta} \ln{Z}= E_0 - 
\frac{1}{\beta}\ln{(1+ \sum_{n > 0}e^{-\beta(E_n-E_0)})} \\
&=& E_0 -\frac{1}{\beta}(\sum_{n > 0}e^{-\beta(E_n-E_0)} +\cdots) 
\nonumber 
\end{eqnarray}
A neglection of high-lying excitations
is not always allowed. For instance, for the one-dimensional 
Ising model in a longitudinal field the
thermodynamic limit and the limit $T\rightarrow 0$ do not commute. 
This is an alternative description of the fact that 
long-range order is destroyed in one dimension. 

In this paper we propose a cumulant formalism for the calculation of
excitation energies of correlated electronic systems. This method is
based on a perturbational approach, i.e., the Hamiltonian is splitted
into $H_0$ and $H_1$ with $H_0$ being exactly solvable. One starts from 
eigenstates of $H_0$ and includes the effect of $H_1$ by an exponential 
ansatz. Especially for the zero-temperature version of the method this can
be shown to be equivalent to summing a perturbation series to infinite
order. So the present method is well suited for systems which can in 
principle be treated perturbatively, but it can not account for systems
with very large fluctuations, e.g., most one-dimensional spin models.

This paper is organized as follows. In the Sec.~II we shortly 
review the size-consistent ground-state version of the cumulant method.
In Sec.~III the cumulant method for the computation 
of excitation energies is presented. With this general scheme we are 
able to calculate partition functions. 
To demonstrate the applicability of the method we investigate the $t$-$J$
model at weak doping as
a topic of current interest. We consider a two-dimensional model with
anisotropic magnetic exchange and calculate the staggered magnetization within
the antiferromagnetic phase in dependence on temperature and hole 
concentration.
A discussion and concluding remarks are put in the last section.


\section{Review of the zero-temperature version of the cumulant method}

For a better understanding of the cumulant method 
its ground-state version is briefly reviewed. 
For more details see\cite{BeckFul88,BeckWonFul89,BeckBre90,Fulde}.
The method starts from the definition of the function
\begin{eqnarray}
\label{4}
  f\left(\lambda\right)=
	 \ln\langle\phi_{0}|e^{-\lambda (H_{0}+H_{1})}
	    e^{\lambda H_{0}}|\phi_{0}\rangle
\end{eqnarray}
where $H_{0}$ is the rigorously solvable unperturbed Hamiltonian $H_0$
with the ground state $|\phi_0\rangle$, i.e., $H_{0}|\phi_{0}\rangle
=\epsilon_{0}|\phi_{0}\rangle$. 
The aim is to calculate the ground-state
energy $E_0$ of the full system, $H |\psi_0\rangle = E_0 |\psi_0\rangle$.
The shift of the ground-state energy $\delta E_{0} = E_0 - \epsilon_0$ 
due to the perturbation $H_{1}$
can be derived in a straightforward way. Introducing the Liouvillian $L_{0}$
which is defined by $L_{0} A = \left[ L_{0} , A \right]$ for any
operator $A$, equation (\ref{4}) is transformed into
\begin{eqnarray}
\label{5}
  f\left(\lambda\right)= \ln\langle\phi_{0}|
                            e^{-\lambda (H_{1}+L_{0})}
		   |\phi_{0}\rangle\,\,\,.
\end{eqnarray}
Next, we define the Laplace transform of the function $f\left(\lambda\right)$
by:
\begin{eqnarray}
\label{6}
  \hat{f}\left(z\right)=-\int\limits_{0}^{+\infty}\,f\left(\lambda\right)\,
					     e^{\lambda\,z}\,dz
  \,\,\,,\hspace{2cm} \Re\{z\} < 0 \,\,\,.
\end{eqnarray}
One can show\cite{BeckFul88,Fulde} that the energy shift $\delta E_{0}$ with 
respect to the unperturbed ground-state energy $\epsilon_{0}$ is given by
\begin{eqnarray}
\label{7}
  \delta E_{0}=\lim_{z\rightarrow 0} z^{2} \hat{f}\left( z \right)\,\,\,.
\end{eqnarray}
On the other side, equation (\ref{5}) is used to express $\delta E_{0}$ 
in terms of cumulants
\begin{eqnarray}
\label{8}
  \delta E_{0}=\lim_{z\rightarrow 0}\,\,
               \langle\phi_{0}|H_{1}\left(1+\frac{1}{z-H_{1}-L_{0}}H_{1}\right)
	                      |\phi_{0}\rangle^{c}\,\,\,.
\end{eqnarray}
Here $\langle\phi_0|...|\phi_0\rangle^c$ denotes cumulant expection 
values with respect to the unperturbed ground state $|\phi_0\rangle$.
Cumulant expectation values\cite{Kubo} for
a product of arbitrary operators $A_i$ with an arbitrary state $|\phi\rangle$
are defined by:
\begin{equation}
\langle\phi|\prod_i A_i^{n_i}|\phi\rangle^c\, = \,
  \left(\prod_i \left({\partial \over \partial\lambda_i} \right)^{n_i} \right)
  \ln\langle\phi|\prod_i {\rm e}^{\lambda_i A_i}|\phi\rangle\,
  |_{\lambda_i=0\,\forall\,i}\,.
\label{CUM_DEF}
\end{equation}
For a note on generalized cumulants see Appendix A.
The quantity inside the bracket of (\ref{8}) is called wave operator $\Omega$
(it has similarity with the M\"oller operator known from scattering theory),
\begin{eqnarray}
\label{10}
\Omega &=&
        1 + \lim_{z\rightarrow 0} \frac{1}{z-H_1-L_{0}}H_{1}
\,.
\end{eqnarray}
Thus we can rewrite $\delta E_{0}$  as
\begin{eqnarray}
\label{11}
  \delta E_{0} &=& \langle\phi_{0}| H_{1} \Omega|\phi_{0}\rangle^{c} 
\ \ \ \ \ \mbox{or} \ \ \ 
E_0 = \langle \phi_0|H \Omega | \phi_0 \rangle^c
\end{eqnarray}
Within cumulants, the operator $\Omega$ transforms the
ground state $|\phi_0\rangle$ of the unperturbed Hamiltonian $H_0$
into the full ground state $|\psi_0\rangle$ of $H$. 
Expanding (\ref{10}) into powers of $H_1$ it can be shown that (\ref{11})
is equivalent to Rayleigh-Schr\"{o}dinger perturbation theory summed
up to infinite order, see e.g. \cite{Fulde}.

There is no general rule how to split $H$ into $H_0$ and $H_1$
except that the overlap between the unperturbed and the full 
ground state has to be non-zero, i.e., 
$\langle\psi_0|\phi_0\rangle\neq 0$.
The operator $\Omega$ describes the influence of $H_1$ onto $|\phi_0\rangle$.
This effect should be a small correction to $|\phi_0\rangle$, i.e.,
it should be treatable perturbatively in the sense that it can be
obtained by summation of a perturbation expansion.
This is usually fulfilled if $H_0$ is the dominant part of the Hamiltonian.
Therefore, for strongly correlated systems $H_0$ should consist 
of the correlations (or at least part of them) whereas $H_1$ 
usually contains the hybridization.
In the application presented in Sec. IV dealing with the weakly doped
$t-J$ model $H_0$ contains the Ising part of the magnetic interaction
whereas $H_1$ consists of its transverse part and the electron hopping.
Other interesting examples may be found in refs. \cite{Polatsek,Koehler}.
If the full ground state $|\psi_0\rangle$ is expected to break a symmetry 
of $H$ then $H_0$ has to be chosen so that its ground state $|\phi_0\rangle$
breaks this symmetry, too.

Instead of using the explicit form (\ref{10}) of the wave operator $\Omega$
an exponential ansatz was proposed \cite{SchorkFul92} 
\begin{eqnarray}
\label{12}
  \Omega = e^S\,,\quad
  S = \sum_{\mu}\alpha_{\mu} S_{\mu}
\end{eqnarray}
where $\{S_{\mu}\}$ is a set of relevant operators. They have to be chosen in
such a way that $\exp(\sum_\mu\alpha_\mu S_\mu) |\phi_0\rangle$
(with appropriate parameters $\alpha_\mu$) represents a good approximation
of the exact ground state.
The yet unknown parameters $\alpha_{\mu}$ are to be determined 
from the following set of equations 
\begin{eqnarray}
\label{13}
  \langle\phi_{0}|S_{\nu}^{\dagger} H \Omega |\phi_{0}\rangle^{c}=0 \,, \,\,
\ \ \ \ \ \nu=1, 2, 3, \cdots
\end{eqnarray}
These equations follow from the condition of $\Omega |\phi_{0}\rangle$ being
an eigenstate of $H$.
Note that equations (\ref{11}) and (\ref{13}) allow for the
computation of the ground-state energy.
For algebraic reasons it is suitable to use operators $S_\mu$ which 
only create (but do not destroy) fluctuations with
respect to the unperturbed ground state $|\phi_0\rangle$. Then the expansion 
of the exponential in (\ref{11}) and (\ref{13}) stops after a few terms
because the Hamiltonian $H$ and the adjoint fluctuation 
operators $S_{\nu}^{\dagger}$ have to remove all fluctuations which
are created by $\Omega$ and $H_1$ in the state $|\phi_0\rangle$.

The choice of appropriate operators $S_\nu$ is most important for
actual calculations using the cumulant method. These operators describe
fluctuations introduced into $|\phi_0\rangle$ by successive application 
of $H_1$.
In principle, they can be derived systematically from the explicit form
(\ref{10}) of the wave operator $\Omega$.
For practical applications this might be only of little help especially
if the main physical effect comes from higher powers of $H_1$.
In such a case a small set of few relevant operators leads to a far simpler
description of the main effect than including a large set of powers 
$H_1^n$. 
The selection of relevant operators for the cumulant method can be seen
similar to the selection of dynamical variables for Mori-Zwanzig
projection technique: Formally, variables can be systematically derived 
from the Liouvillian, but often choosing variables from physical
insight is more useful.

For a discussion of the cumulant method and its relation to other
methods like variational calculations and the coupled-cluster method see
appendix C.
In the past, equations (\ref{11},\ref{13}) was used to evaluate
ground-state properties of several systems, see for instance
\cite{BeckWonFul89,Polatsek,Koehler,BEW,VojBeck1,VojBeck2}.


\section{Extension to finite temperatures}

Now we present a cumulant scheme for calculating excitation energies. 
To discuss the influence of the perturbation $H_1$ we formally introduce
an additional parameter $\lambda$ in the Hamiltonian:
\begin{equation}
H\;=\;H_0\,+\,\lambda H_1
\end{equation}
As is shown below, the wave operator $\Omega$
from the zero-temperature approach will be replaced by a unitary operator $U$ which
diagonalizes the full Hamiltonian $H$. 
The operator $U$ transforms all eigenstates of the unperturbed system into
eigenstates of the full Hamiltonian. Being $|\phi_n\rangle$ and 
$|\psi_n\rangle$ the eigenstates of $H_0$ and $H$,
\begin{equation}
\label{14a}
H_0 |\phi_n\rangle = \epsilon_n|\phi_n\rangle
\ \ , \ \ \ \ \ \ \ \ \
H|\psi_n\rangle = E_n |\psi_n\rangle
\, ,
\end{equation}
the action of $U$ is defined by
\begin{equation}
\label{14b}
         |\psi_n\rangle = U |\phi_n\rangle
\end{equation}
with $U$ depending smoothly on $\lambda$ and $U \rightarrow 1$ 
for $\lambda \rightarrow 0$.
Every unitary operator can be written as
\begin{equation}
U \:=\: e^S \quad \mbox{with} \quad
S^\dagger \,=\, -S
\,.
\end{equation}

In general, the unitary transformation $U$ is not known 
for a given system. Therefore approximations for $U$
have to be used. Generalizing the exponential form (\ref{12})
for the wave operator $\Omega$ from the zero-temperature approach
to a unitary operator, we make the following ansatz 
\begin{eqnarray}
\label{14c}
  U = e^S\,,\quad 
  S = \sum_{\mu}\alpha_{\mu}(S_{\mu}-S_{\mu}^{\dagger})
\end{eqnarray}
where both $S_{\mu}$ and the adjoint operators
$S_{\mu}^\dagger$ are enclosed in the exponential.
The $\{S_\nu\}$ form a set of relevant operators as in the zero-temperature
approach. In most cases, using a finite number of operators $S_\nu$ 
represents an approximation of $U$. However, with the number of operators
going to infinity the exact transformation $U$ can be approximated with
arbitrary accuracy.

In contrast to the ground-state approach the operators $S_{\mu}$ and 
$S_{\mu}^\dagger$ can create and also annihilate fluctuations.
This also means that the application of $(S_{\mu}-S_{\mu}^{\dagger})$
to an eigenstate $|\phi_n\rangle$ of $H_0$ (with energy $\epsilon_n$)
leads to a new state which may overlap with
eigenstates of $H_0$ having energies both below and
above $\epsilon_n$.
 
In the finite-temperature approach the former equations 
(\ref{11}) and (\ref{13}) are replaced by
\begin{eqnarray}
\label{15}
  E_n \:&=&\:  \langle H U \rangle_n^c  \,, \\
\label{16}
   0  \:&=&\:  \langle (S_{\nu} - S_{\nu}^{\dagger})^{\dagger}
               H U \rangle_n^c \,,
   \ \ \ \ \ \nu=1, 2, 3, \cdots
\end{eqnarray}
where $\langle \cdots \rangle_n^c$ refers to the cumulant 
average now formed with the unperturbed eigenstate $|\phi_n\rangle$
of $H_0$, i.e., $\langle \cdots \rangle_n^c = 
\langle\phi_n| \dots |\phi_n\rangle^c$. The parameters $\alpha_{\mu}$
can be determined from Eqs.~(\ref{16}). Note that the cumulants in
(\ref{16}) can be formed with any unperturbed eigenstate 
$|\phi_n\rangle$ of $H_0$ as long as the exact expression for 
the unitary transformation $U$ is known. However, for any approximation
for $U$ the parameters $\alpha_{\mu}$ may vary 
with a different choice of the $|\phi_n\rangle$ used for the 
cumulant expectation values.

To prove Eqs. (\ref{15},\ref{16}) one can either make use of a
method recently proposed \cite{Kladko} which 
is based on integrating over infinitesimal transformations $(1+S/N)$. 
Here we rather prefer to make explicit use of 
the definition (\ref{CUM_DEF}) of cumulant expectation values.
Expanding $U= e^S$ in powers of $S$ and taking the 
definition (\ref{CUM_DEF})
for cumulant expectation values one finds for the r.h.s. of (\ref{15})
\begin{eqnarray}
\label{17}
  \langle H e^S \rangle_{n}^{c}&=&
\sum_m^{\infty} \frac{1}{m!} \langle H S^m \rangle_n^c
\\
&=&  \sum_{m}^{\infty}\frac{1}{m!}
		   \lim_{\lambda_{1},\lambda_{2}\rightarrow 0}\,
  \frac{\partial}{\partial \lambda_{1}}\,
  \frac{\partial^{m}}{\partial \lambda_{2}^{m}}\,
  \ln\langle\phi_{n}\vert 
		    e^{\lambda_{1} H} \
		    e^{\lambda_{2} S}
		    \vert\phi_{n}\rangle             \nonumber
\,.
\end{eqnarray}
Summing over $m$ one immediately finds the desired result
\begin{eqnarray}
\label{18}
\langle H e^{S} \rangle_n^c &=&
       \left .
       \frac{\partial}{\partial \lambda_1} \ln{ 
        \langle \phi_n| e^{\lambda_1H}\ e^{(\lambda_2 + 1) S}
           |\phi_n\rangle}
        \right |_{\lambda_1=\lambda_2=0} \\
       &=& \left .
           \frac{\partial}{\partial \lambda_1} \ln{ \langle \phi_n|
           e^{\lambda_1 H} \ e^{S} |\phi_n \rangle} 
         \right |_{\lambda_1=0} \nonumber \\
       &=& \left .
            \frac{\partial}{\partial \lambda_1} 
            \ln{\langle \phi_n| e^{\lambda_1 H}| \psi_n\rangle}
            \right |_{\lambda_1=0}
            \,=\, E_n
\,.
\nonumber
\end{eqnarray}
The second equation (\ref{16}) can be proved in close analogy.

The changes imposed by the replacement of $\Omega$ by $U$
give rise to some problems for actual calculations.
The expansion of $U$ does not terminate any longer
after a few steps because the powers of $S_\mu^\dagger$ in the unitary
operator $U$ can remove fluctuations created before by $S_\mu$.
For this reason an infinite series occurs. It may be possible, however,
to find a solution in two steps:
(i) First, the cumulants can be rewritten in terms
of usual expection values formed with $|\phi_n\rangle$ (see Appendix B),
e.g.
\begin{eqnarray}
\label{19}
  0&=& \langle S_\nu H e^{S} \rangle_{n}^{c} =
  \frac{\langle S_\nu H e^{S}\rangle_{n}}
  {\langle e^{S}\rangle_{n}}- 
  \frac{ \langle S_\nu e^{S} \rangle_{n}}
       {\langle e^{S} \rangle_n}
  \frac{\langle H e^{S} \rangle_{n}}
       {\langle e^{S} \rangle_n}  \\
&& \nonumber \\
\label{20}
  E_n &=& \langle H e^{S} \rangle_{n}^{c} =
  \frac{ \langle H e^{S} \rangle_{n}}
       { \langle e^{S} \rangle_{n}}\,\,\,,
\end{eqnarray}
with $U = e^S$ and $S = \sum_{\mu}\alpha_{\mu}(S_\mu - S_\mu^\dagger)$.
These relations can be easily verified
by use of equation (\ref{CUM_DEF}).
(ii) In a second step the exponential
$e^{S}$ should be decomposed. The basic idea is 
to extract a factor next to $|\phi_n \rangle$ that only  
annihilates fluctuations when it is applied to $|\phi_n\rangle$. 
This means that the power series of this factor 
usually stops after a few terms.
As an example we consider the factorization for the special case
of $S_{\mu}$ being a boson operator, 
$[S_{\mu}, S_{\mu}^{\dagger}]= 1$. In this case a well known decomposition for 
$U$ reads 
\begin{eqnarray}
\label{21}
  U &=& e^{\alpha (S_{\mu}-S_{\mu}^{\dagger})}=
  e^{-\alpha S_{\mu}^{\dagger}}\ 
  e^{-\frac{\alpha^2}{2}} \ 
  e^{\alpha S_{\mu}}\,\,\,.
\end{eqnarray}
If $U$ is applied to an eigenstate of $H_0$ containing a finite number $n$
of bosons, $(S_{\mu}^{\dagger})^n$, the power series of the factor
$e^{\alpha S_{\mu}}$ stops after $n+1$ steps. For $n=0$ (ground state)
only the identity operator would survive.
Proceeding this way the application of $U$ is much easier to 
evaluate than before. 
In general, decompositions like (\ref{21}) depend on the details of the 
given problem.
The example in Sec. IV will show that the outlined evaluation scheme is an 
appropriate tool to account for the influence of low-lying excitations
especially for low temperatures.

If we are faced with degenerate eigenstates of $H_0$ with a
degeneracy lifted by $H_1$ we have the freedom to choose initial states
$|\phi_n\rangle$ for the cumulant method (\ref{15},\ref{16}) from the subspace of
degenerate states. In principle for every choice of $|\phi_n\rangle$ a unitary
transformation diagonalizing the Hamiltonian can be found. But if we use a
certain ansatz for the operator $U$ then the best choice of $|\phi_n\rangle$
depends on the form of this ansatz. Assume that the states
$|\phi_{n_1}\rangle, ..., |\phi_{n_m}\rangle$ are degenerate eigenstates of
$H_0$ with energy $\epsilon_n$. The correct states
$|{\tilde\phi}_{n_i}\rangle$ with $U |{\tilde\phi}_{n_i}\rangle$ being
eigenstates of $H$ (for a certain form of $U$)
are linear combinations of the $|\phi_{n_i}\rangle$:
\begin{equation}
|{\tilde\phi}_{n_i}\rangle = \sum_{j=1}^m \gamma_{ij} |\phi_{n_j}\rangle
\,.
\label{LINKOMB}
\end{equation}
To find the $\gamma_{ij}$ one uses again Eq. (\ref{16})
provided by the cumulant method. Note that Eq. (\ref{16}) also holds for generalized
cumulants\cite{Kladko} (see Appendix A), i.e., defined with a bra vector different
from the ket $|\phi_n\rangle$:
\begin{equation}
   0  =  \langle\Phi| (S_{\nu} - S_{\nu}^{\dagger})^{\dagger}
               H U |\phi_n\rangle^c
\,.
\end{equation}
The arbitrary vector $\langle\Phi|$ should have a non-zero overlap with
$|\phi_n\rangle$.
Evaluating the cumulants in analogy to (\ref{19},\ref{20}) 
and using the linear combination (\ref{LINKOMB}) instead of
the ket vector $|\phi_n\rangle$ we obtain
\begin{equation}
\sum_j \gamma_{ij} 
\langle \Phi | S_\nu H U | \phi_{n_j} \rangle
= E_{n_i}
\sum_j \gamma_{ij} 
\langle \Phi | S_\nu U |\phi_{n_j} \rangle
\,.
\label{DEG_EVAL}
\end{equation}
For fixed $U=e^S$ and appropriate operators $S_\nu$, Eq.
(\ref{DEG_EVAL}) is a generalized eigenvalue problem for the $E_{n_i}$
and $\gamma_{ij}$ and can be solved by standard methods.

We should like to mention that the method presented here for calculating excitation
energies on the basis of cumulants is rather different from the method
proposed recently by Schork and Fulde\cite{SchorkFul94}. There 
the zero-temperature wave operator $\Omega$ (\ref{10}) was applied to excited
eigenstates $|\phi_n\rangle$ of $H_0$ to obtain the full eigenstate
$|\psi_n\rangle$. This approach is only valid if
$|\langle\psi_n|\phi_n\rangle|^2 \ge \frac {1} {2}$ which is a significant
restriction compared to the present approach based on the introduction of the
transformation $U$.


\section{Application to the $t$-$J$ model at weak doping}

The thermodynamics of the antiferromagnetic phase in high-temperature 
superconductors is discussed as the major application of the present 
cumulant approach for finite temperatures.
It is widely accepted that the electronic properties of these
systems are mainly determined by the charge carriers in the 
CuO$_2$ planes. Of special interest is the 
interplay between antiferromagnetism and hole doping. Neutron scattering 
experiments\cite{Shirane,RossMig91} show that the effective magnetization
strongly depends on the hole concentration $\delta$ within 
the CuO$_{2}$ planes. Both the N\'{e}el temperature and 
the staggered magnetization decrease
rapidly with increasing $\delta$ and vanish at a critical 
concentration $\delta_{c}$. For reviews on experimental and theoretical
investigations of high-T$_c$ superconductors see refs.
\cite{Dagotto94,Brenig95}.

The essential aspects of the low-energetic
electronic degrees of freedom of the $\rm CuO_2$ planes
are by now believed to be well described by the two-dimensional $t$-$J$
model\cite{Anderson87,Zhang88}:
\begin{equation}
H\, =\, - t \sum_{\langle ij\rangle \sigma}
      (\hat c^\dagger_{i\sigma} \hat c_{j\sigma} +
       \hat c^\dagger_{j\sigma} \hat c_{i\sigma})
    + J \sum_{\langle ij\rangle} \ ({\bf S}_i {\bf S}_j - {{n_i n_j} \over 4} )
\label{TJ_MOD}
\end{equation}
As above, ${\bf S}_i$ is the electronic spin operator and $n_i$ the electron
number operator at site $i$.
Note that the Hamiltonian (\ref{TJ_MOD}) is defined in the subspace of the
unitary space without double occupations of sites. The
electronic creation operators $\hat c_{i \sigma}^{\dagger}$ are not usual
fermion operators but rather exclude double occupancies:
\begin{equation}
\hat c^{\dagger}_{i\sigma} = c^{\dagger}_{i\sigma} (1-n_{i,-\sigma})
\end{equation}
At half filling the $t$-$J$ Hamiltonian reduces to the antiferromagnetic
Heisenberg model.

In a recent paper\cite{VojBeck2} we have investigated the behavior
of the ground-state
sublattice magnetization with increasing hole concentration
using the cumulant formalism for zero temperature. Here we want to generalize
these results and calculate the temperature and doping dependence of the 
magnetization. 
We consider the situation of weak doping, i.e., of a few
non-interacting holes present in the half-filled system. 
As done in ref.\cite{VojBeck2} we will use approximations equivalent
to linear spin-wave theory for expectation values with spin operators
and assume independent hole motion.
Several analytical
calculations at zero\cite{VojBeck2,IgFul92,KhaHo,BelRi} and finite temperature
\cite{RiYush} show that the strong coupling between holes and spin waves leads
to a softening of long-wavelength spin excitations. This softening causes
a reduction of the order parameter at zero temperature as well as a decrease
of the N\'{e}el temperature with doping.

Note that a non-zero spontaneous magnetization in a
two-dimensional isotropic system at finite temperature is strictly
forbidden by the Mermin-Wagner theorem\cite{MerminWagner}.
The high-$T_c$ materials show a spontaneous magnetization at $T>0$ because
of the anisotropy due to the magnetic coupling between the copper-oxide layers.
For a simple description of that fact we introduce an 
anisotropy in the Heisenberg exchange, 
i.e., we use an exchange constant of $J(1+\epsilon)$ in
$z$ direction. Within linear spin-wave theory this is
equivalent to an additional staggered field $\epsilon J$ parallel to the
$z$-axis. The value of the anisotropy $\epsilon$ can be
fitted from the N\'{e}el temperature $T_N$ observed in experiments.
It can also be extracted from NMR data, see e.g. \cite{RossMig91,Bucci93}.
Experimental values are $J\approx 0.15 eV$ and $T_N\approx 400 K\approx 0.25J$ 
in the undoped materials (YBCO)\cite{RossMig91}, so we are interested in
the temperature range $0...0.5 J$.

\subsection{Undoped case}

First we briefly consider the half-filled case, i.e., the 
undoped system.
The $t$-$J$ model reduces to the $s=1/2$ Heisenberg antiferromagnet.
For the application of the cumulant formalism we decompose the anisotropic 
Heisenberg Hamiltonian as follows:
\begin{eqnarray}
H_{0}\, &=&\,\, H_{Ising} \nonumber\\
&=&\,\, J\,(1+\epsilon)\,
  \sum_{<ij>}(S_{i}^{z}\, S_{j}^{z}\,-\, \frac{n_{i} n_{j}} {4}\, ) \, ,
\label{HHEIS}  \\
H_{1}\, &=&\,\, H_{\bot} 
  \,\,=\,\, 
  \frac{J} {2}\,\sum_{<ij>}(\, S_{i}^{-}S_{j}^{+}\,+\, S_{i}^{+}S_{j}^{-}
  \, )\, \nonumber.
\end{eqnarray}
We will use linear spin-wave approximation for the spin operators to calculate
expectation values, i.e., the Fourier-transformed spin operators 
$S_{\bf q}^\pm$ are replaced by
bosons $a_{\bf q}$ and $b_{\bf q}$.
These operators 
are spin-wave operators for
the $\uparrow$-and $\downarrow$-sublattice defined by
\begin{eqnarray}
a_{\bf q}^+ \,\,&=&\,\, \frac{1} {\sqrt{N/2}}\quad
  \sum_{i\in\uparrow}e^{i{\bf q}{\bf R}_i}
  \, S_{i}^{-}\,\, , \label{AQ_DEF} \\
b_{\bf q}^+ \,\,&=&\,\, \frac{1} {\sqrt{N/2}}\quad
  \sum_{j\in\downarrow}e^{i{\bf q}{\bf R}_j}
  \, S_{j}^{+}\,\, . \nonumber
\end{eqnarray}
In the approximation of linear spin-wave theory they 
obey usual Bose commutation relations:
\begin{equation}
[a_{{\bf q}_1},a^+_{{\bf q}_2}]_- \,=\,\delta_{{\bf q}_1 {\bf q}_2}\, , \quad
[b_{{\bf q}_1},b^+_{{\bf q}_2}]_- \,=\,\delta_{{\bf q}_1 {\bf q}_2}\, , \quad
[a_{{\bf q}_1}^{(+)},b_{{\bf q}_2}^{(+)}]_- \,=\,0 .
\end{equation}

Within linear spin-wave approximation the Hamiltonian
(\ref{HHEIS}) takes the form
\begin{eqnarray}
H_{0}\,\, &=&\,\, {J\,(1+\epsilon)\,z_0 \over 2}
  \left ({N \over 4} \,\,-\,\,
  \sum_{\bf q} \left (
  a_{\bf q}^+a_{\bf q} \,+\,b_{\bf q}^+b_{\bf q}
  \right ) \,
  \right ) \, , \nonumber\\
H_{1}\,\, &=& \,\,
  \frac {J z_0} {2}\,\sum_{\bf q} \gamma{(\bf q)} \left (
  a_{\bf q} b_{-\bf q} \,+\,a_{\bf q}^+b_{-\bf q}^+
  \right )
\label{HHEISSPW}
\end{eqnarray}
where $z_0\gamma({\bf q}) \,=\, 2(\cos q_x\,+\,\cos q_y)$ is the
Fourier-transformed exchange coupling and $z_0$ the number of nearest
neighbor sites.
The wave vectors $\bf q$ have to be taken from the magnetic Brillouin zone.

The ground state $|\phi_0\rangle$ of the unperturbed Hamiltonian $H_0$ is the
antiferromagnetically ordered N\'{e}el state. The excited states of $H_0$
are obtained by creating fixed numbers of bosons $a_{\bf q}^+$ and $b_{\bf q}^+$
in the N\'{e}el state:
\begin{equation}
|\phi_{\{n_{a\bf q},n_{b\bf q}\}}\rangle \quad=\quad
\prod_{\bf q} \,\,
  { (a_{\bf q}^+)^{n_{a\bf q}} \over \sqrt{n_{a\bf q}!} } \,
  { (b_{\bf q}^+)^{n_{b\bf q}} \over \sqrt{n_{b\bf q}!} } \,\,
|\phi_{N{\mathaccent 19 e}el}\rangle
\label{HEIS_EXSTATES}
\end{equation}
where $\{n_{a\bf q},n_{b\bf q}\}$ is the set of numbers of spin-wave
excitations with momenta $\bf q$ on the
$\uparrow$-and $\downarrow$-sublattices.

In the unitary operator $U$ of the cumulant method we have to include 
the effect of the perturbation $H_1$ which can either create or destroy pairs
of spin waves. An appropriate form of $U$ is
\begin{equation}
U = \prod_{\bf q} U_{\bf q} \, , \quad
U_{\bf q}\,\, =\,\,
  e^{ \mu_{\bf q}\,
  (a_{\bf q}^+ b_{-\bf q}^+ \, - \, 
   a_{\bf q} b_{-\bf q}) }
\,. \label{UHEIS1}
\end{equation}
Note that the operators $U_{\bf q}$ commute with each other because 
$a_{\bf q}$ and $b_{\bf q}$ are independent bosons (linear spin-wave 
approximation).
The coefficients $\mu_{\bf q}$ have to be determined by solving the equation
$0\,=\,\langle\phi_n| A\,H\,U |\phi_n\rangle^c$ for appropriate operators $A$
and eigenstates $|\phi_n\rangle$ of $H_0$. After transforming
the cumulants into normal expectation values according to Appendix B
we have to apply $U$ to the eigenstates $|\phi_{\{n_{a\bf q},n_{b\bf q}\}}\rangle$
of $H_0$.
In Appendix D it is shown that $U_{\bf q}$ can be factorized into
the following form:
\begin{eqnarray}
U_{\bf q}\,\,=&&\,\,
  e^{ \displaystyle (\tanh \mu_{\bf q}) \, a_{\bf q}^+ b_{-\bf q}^+ } \,
  e^{ \displaystyle -\, (\ln\, \cosh \mu_{\bf q}) \, 
                        (a_{\bf q}^+a_{\bf q}+b_{-\bf q}^+b_{-\bf q}+1) } \,
  e^{ \displaystyle -\, (\tanh \mu_{\bf q})\,a_{\bf q} b_{-\bf q} }
  \,. \label{UHEIS2}
\end{eqnarray}
If one applies $U$ to the unperturbed ground state $|\phi_0\rangle$
one finds
\begin{eqnarray}
U\,|\phi_0\rangle
\,\,=\,\, \left (\prod_{\bf q}\, \sqrt {1-\nu_{\bf q}^2} \right) \,\, 
  e^ { \sum_{\bf q}\, \nu_{\bf q} \, a_{\bf q}^+ b_{-\bf q}^+ } \,
|\phi_0\rangle \,.
\end{eqnarray}
with the substitution $\nu_{\bf q}\,=\,\tanh \mu_{\bf q}$.
This transformation is the same as used within the ground-state cumulant
formalism to describe the Heisenberg antiferromagnet, see\cite{VojBeck1}.
It can be used to determine the coefficients $\nu_{\bf q}$.
From
\begin{equation}
0 \,\,=\,\, \langle\phi_0| (a_{\bf q} b_{-\bf q})^{\cdot}\, H U 
           |\phi_0\rangle^c
\end{equation}
we obtain a quadratic equation for each $\nu_{\bf q}$.
It has the solution\cite{VojBeck1}
\begin{equation}
\tanh \mu_{\bf q} \,\,=\,\,
\nu_{\bf q} \,\, = \,\, \frac {1} {\gamma({\bf q})}
  \left (
  - (1+\epsilon) \,+\,\sqrt{ (1+\epsilon)^2\,-\,\gamma({\bf q})^2 }
  \right )
\, .
\end{equation}
With these values for the coefficients the unitary transformation $U$ 
(\ref{UHEIS1})
equals the usual Bogoliubov transformation for the antiferromagnet
which diagonalizes the Hamiltonian (\ref{HHEISSPW}).
Therefore we expect to obtain exactly the results known from
linear spin-wave theory.

In order to calculate the energy spectrum of the system we have to evaluate
\begin{equation}
E_{\{n_{a\bf q},n_{b\bf q}\}} \,\,=\,\,
\langle\phi_{\{n_{a\bf q},n_{b\bf q}\}}| H\, U
|\phi_{\{n_{a\bf q},n_{b\bf q}\}}\rangle^c \,\,=\,\,
{ \langle\phi_{\{n_{a\bf q},n_{b\bf q}\}}| H\, U
  |\phi_{\{n_{a\bf q},n_{b\bf q}\}}\rangle \over
  \langle\phi_{\{n_{a\bf q},n_{b\bf q}\}}| U
  |\phi_{\{n_{a\bf q},n_{b\bf q}\}}\rangle }
\,.
\label{E_N_HEIS}
\end{equation}
This is done in Appendix E. It leads to the following result:
\begin{eqnarray}
E_{\{n_{a\bf q},n_{b\bf q}\}} \,\,&=&\,\,
  E_{N{\mathaccent 19 e}el} \,+\,
  2J \sum_{\bf q} (n_{a\bf q}+n_{b\bf q}) \,+\,
  \sum_{\bf q} \frac J 2 \gamma({\bf q}) \nu_{\bf q} \,
  (n_{a\bf q}+n_{b\bf q}+1) \nonumber\\
&=&\,\,E_0 \,+\, \sum_{\bf q} (n_{a\bf q}+n_{b\bf q})\, \omega_{\bf q}
\label{EHEIS2}
\end{eqnarray}
where $\omega_{\bf q}$ is the spin-wave energy given by
\begin{equation}
\omega_{\bf q}
   \,\,=\,\, {Jz_0 \over 2}\,
             \left(1+\epsilon\,+\, \gamma({\bf q}) \nu_{\bf q} \right)
   \,\,=\,\, {Jz_0 \over 2}\, 
             \sqrt{(1+\epsilon)^2\,-\,\gamma({\bf q})^2}
\,\,.
\end{equation}
$E_0$ denotes the ground-state energy of the Heisenberg antiferromagnet 
within linear spin-wave approximation.
Eq. (\ref{EHEIS2}) exactly equals the result of linear spin-wave
theory. It describes the spectrum of a system of independent
bosons with the dispersion $\omega_{\bf q}$.

\subsection{Weakly doped case}

Now we turn to the discussion of the weakly doped system.
The $t$-$J$ Hamiltonian with anisotropic magnetic exchange can be decomposed
as follows:
\begin{eqnarray}
H_{0}\, &=&\,\, H_{Ising}\,+\,H_{Zeeman} \nonumber\\
&=&\,\, J\,(1+\epsilon)\,
  \sum_{<ij>}(S_{i}^{z}\, S_{j}^{z}\,-\, \frac{n_{i} n_{j}} {4}\, )
   \,+\, g_J \mu_B B_A\,(-\sum_{i\in\uparrow} S_i^z\,+\,
                          \sum_{j\in\downarrow} S_j^z ) ,
\label{H_TJ}  \\
H_{1}\, &=&\,\, H_{t}\, +\, H_{\bot} \nonumber\\
&=&\,\, -t\, \,\sum_{<ij>,\sigma}(\,\hat{c}_{i\sigma}^{+}\,
  \hat{c}_{j\sigma}\, \,+\, \,\hat{c}_{j\sigma}^{+}\,\hat{c}_{i\sigma}\,)
  \,+\, \frac{J} {2}\,\sum_{<ij>}(\, S_{i}^{-}S_{j}^{+}\,+\, S_{i}^{+}S_{j}^{-}
  \, )\, \nonumber.
\end{eqnarray}
We have added a staggered field $B_A$ in $z$-direction which will later be used
to determine the magnetization of the system. 

The ground state $|\phi_0\rangle$ of $H_0$ is a N\'{e}el state with 
$M=\delta\cdot N$ holes where $\delta$ is the hole concentration and $N$ the
number of lattice sites. The holes have fixed momenta ${\bf k}_m$ and are 
located on the sublattice $\sigma_m = \uparrow,\downarrow$.
Excited eigenstates of $H_0$ can be defined by
(compare (\ref{HEIS_EXSTATES}) ):
\begin{eqnarray}
|\phi_{\{{\bf k}\sigma\},\{n_{a\bf q},n_{b\bf q}\}}\rangle
 &=&\quad
\prod_{\bf q} \,\,
  { (a_{\bf q}^+)^{n_{a\bf q}} \over \sqrt{n_{a\bf q}!} } \,
  { (b_{\bf q}^+)^{n_{b\bf q}} \over \sqrt{n_{b\bf q}!} } \,\,
\prod_{m=1}^M \hat{c}_{{\bf k}_{m}\sigma_{m}}
  \, |\phi_{N{\mathaccent 19 e}el}\rangle
\label{GSDEF4}
\end{eqnarray}
The set of hole momenta (and spins) in this state is denoted by
$\{{\bf k}\sigma\}$. Their distribution can be different from the
one of the ground state. The product $\prod_m$ runs over all 
holes in the system $({\bf k}_m\sigma_m\,\in\,\{{\bf k}\sigma\})$, and
$\{n_{a\bf q},n_{b\bf q}\}$ is the set of numbers of spin-wave
excitations with momenta $\bf q$ on the $\uparrow$-and $\downarrow$-sublattices.

For a proper description of the hole states we use path (or string) states
\cite{BrinkRice,Nagaoka,Trugman88,ShrSig88} which lead
to the concept of spin-bag quasiparticles.
The path states are generated by repeated application of the hopping
Hamiltonian $H_t$ to a hole in a N\'{e}el background. The moving
hole leaves behind a trace of overturned spins. 
Here we define path concatenation operators $A_n=A_{n\uparrow}+
A_{n\downarrow}$. The operators
$A_{n\uparrow}$ and $A_{n\downarrow}$ refer to the two sublattices.
$A_{n\uparrow}$ operating on the N\'{e}el state with one hole,
$\hat c_{i\uparrow} |\phi_{N{\mathaccent 19 e}el}\rangle$, moves the hole
$n$ steps away and creates a path or string of $n$ spin defects
attached to the transferred hole. Explicitly, the operators $A_{n\uparrow}$
are defined by
\begin{eqnarray}
A_{1\uparrow}\quad &=&\quad \frac{-1} {\sqrt{z_{0}}}\quad\sum_{ij}\hat{c}_{j\downarrow}
        \hat{c}_{i\downarrow}^{+}\, {\tilde R}_{ji}\, ,\nonumber\\
A_{2\uparrow}\quad &=&\quad \frac{1} {\sqrt{z_{0}\, (z_{0}-1)}}\quad\sum_{ijl}
        \hat{c}_{l\uparrow}S_{j}^{+}\hat{c}_{i\downarrow}^{+}
        \, R_{lj}^{(i)}\, {\tilde R}_{ji}\, , 
\quad\quad\quad (i\in\uparrow,\, j\in\downarrow,\, l\in\uparrow,\, m\in\downarrow)
\label{PATHOPDEF} \\
A_{3\uparrow}\quad &=&\quad \frac{-1} {\sqrt{z_{0}\, (z_{0}-1)^{2}}}\quad\sum_{ijlm}
        \hat{c}_{m\downarrow}S_{l}^{-}S_{j}^{+}\hat{c}_{i\downarrow}^{+}
        \, R_{ml}^{(j)}\, R_{lj}^{(i)}\, {\tilde R}_{ji}\, ,\nonumber\\
\ldots\quad&\quad&\nonumber
\end{eqnarray}
The operators $A_{n\downarrow}$ for the
'down' sublattice are defined analogously with all spins reversed.
$z_0$=4 denotes the number of nearest neighbor
sites in the lattice. The matrices ${\tilde R}_{ji}$
and $R_{lj}^{(i)}$ allow the hole to jump to its four
nearest neighbors in the first step and to only three
new nearest neighbors by hopping forward in each
further step:
\begin{eqnarray}
{\tilde R}_{ji}\quad &=&\quad\left\{
   \begin{array}{rl}
     1&\quad i,\, j\quad {\rm nearest\,neighbors}\\
     0&\quad {\rm otherwise}\end{array}\right.\quad ,\\
R_{lj}^{(i)}\quad &=&\quad\left\{
    \begin{array}{rl}
      1&\quad j,\, l\quad {\rm nearest\,neighbors\,and}\, l\not= i \\
      0&\quad {\rm otherwise}
    \end{array}\right.\quad . \nonumber
\end{eqnarray}

The staggered magnetization per spin and $\mu_B$
($\mu_B$ - Bohr magneton)
has to be obtained from the free energy $F$ by
\begin{equation}
M_{eff} \,\,=\,\,
   {1 \over N}{\rm Tr} \, \rho(\sum_{i \in \uparrow}  S_i^z -
               \sum_{j \in \downarrow}S_j^z )
        \,\,=\,\,
   - \left . {\partial F \over N \mu_B \partial B_A} \,
                \right |_{B_A=0}
\label{MAG_FDIFF}
\end{equation}
where $\rho$ denotes the statistical operator of the full system described
by $H$.

In the doped system we have two possible contributions to the reduction of 
the order parameter: the spin bag, which forms the quasiparticle around each
hole, and spin-wave excitations.
From ground state calculations\cite{VojBeck2,BelRi}, however, it is known 
that the main contribution to the decrease of the magnetization arises from
spin waves which admix into the ground state. We expect that also the decrease 
of the magnetization with increasing temperature comes primarily from
excited spin waves. These spin waves will be treated as bosonic excitations, i.e.,
spin-wave interactions will be neglected. Within our model we shall obtain
a spin-wave spectrum which depends on the number and momentum distribution
of the hole quasiparticles. 

Since the {\rm direct} contribution from each spin-bag quasiparticle to the
decrease of the magnetization is small, we assume it to be
independent of temperature and hole concentration. This means that structure 
and size of the quasiparticle (for fixed momentum) do not change with 
temperature and doping.
Consistent with this we employ a rigid-band approximation for the 
quasiparticles, i.e., we keep the hole dispersion $\epsilon_{\bf k}$
fixed (at its value at $\delta=0$ and $T=0$). In the ground state all holes
are found near the minima of the dispersion being located at
${\bf k}=(\pm\pi/2, \pm\pi/2)$ (Fermi sea). Considering excited hole states we
take into account particle-hole excitations within the lowest quasiparticle
band. However, 
excitations to higher bands (interband excitations) are neglected because such
excitations would have energies of order $2J$ and the temperatures we are
interested in are smaller than $J/2$.

In the unitary operator $U$ of the cumulant method we have to include 
the effect of the perturbation $H_1$. Its part $H_\perp$ creates or destroys pairs
of spin waves. $H_t$ provides hole hopping which can be described via
the path operators (\ref{PATHOPDEF}). Within a factorization
approximation $U$ takes the form
\begin{eqnarray}
U &=& \left(\prod_{\bf q} U_{\bf q}^{(spins)}\right) \,U^{(holes)}\, ,\nonumber\\
U_{\bf q}^{(spins)}\,\, &=&\,\,
  e^{ \mu_{\bf q}\,
  (a_{\bf q}^+ b_{-\bf q}^+ \, - \, 
   a_{\bf q} b_{-\bf q}) } \, ,\nonumber\\
U^{(holes)}\, \, &=&  \,\,
   e^{\sum_{n}\lambda_n A_{n}}
\,. \label{UTJ1}
\end{eqnarray}
This ansatz generalizes the operator $U$ from (\ref{UHEIS1}) to the 
case of weak doping.
In the hole part $U^{(holes)}$ we only include path creation operators because
we consider intraband but not interband hole excitations.
We will see that the (intraband) hole excitations affect the spin-wave spectrum, 
i.e., the spin-wave renormalization depends on the distribution
of the hole momenta in the system. 
Within our factorization approximation we therefore have to use
coefficients $\mu_{\bf q}$ in $U^{(spin)}$ which depend
on the set of hole momenta $\{{\bf k}\sigma\}$. 
Path operators $A_n$ are included up to length $n=2$.
The path coefficients $\lambda_n$ should depend on the hole momenta 
in (\ref{GSDEF4}) and on
the spin-wave excitations. As discussed above, we expect this dependence to be weak.
It is neglected within our rigid-band approximation which fixes
the values of the $\lambda_n$ to those known from the ground-state calculation
\cite{VojBeck2}.

With the ansatz (\ref{UTJ1}) we can now calculate the coefficients $\mu_{\bf q}$
for all distributions of hole momenta. In analogy to the undoped case
we use the equation
\begin{equation}
0 \,\,=\,\, \langle\phi_{\{{\bf k}\sigma\},\{0,0\}}|
           (a_{\bf q} b_{-\bf q})^{\cdot}\, H U 
           |\phi_{\{{\bf k}\sigma\},\{0,0\}}\rangle^c
\label{DOP_QUADGL}           
\end{equation}
which is provided by the cumulant method. The expectation values are
calculated to first order in $\delta$ using linear spin-wave theory, for
details see\cite{VojBeck1,VojBeck2}.

Having found the values for the coefficients $\mu_{\bf q}$ (the $\lambda_n$
are fixed by the rigid-band approximation for the hole quasiparticles)
we can calculate the energies of the system states. These are described by the
set of hole momenta $\{{\bf k}\sigma\}$ and the numbers of spin-wave
excitations $\{n_{a\bf q},n_{b\bf q}\}$.
The energies are calculated in analogy to Appendix D,
they can be written as
\begin{eqnarray}
E_{\{{\bf k}\sigma\},\{n_{a\bf q},n_{b\bf q}\}} \,\,&=&\,\,
  \langle\phi_{\{{\bf k}\sigma\},\{n_{a\bf q},n_{b\bf q}\}}| HU
  |\phi_{\{{\bf k}\sigma\},\{n_{a\bf q},n_{b\bf q}\}}\rangle^c \nonumber\\
 &=& \,\,
    E_{N{\mathaccent 19 e}el}\,(1-\delta)
     \,-\,(2J(1-\delta)+B_A) N \,+\, \nonumber\\
 &+&\,\,
     \sum_m \epsilon_{{\bf k}_m} \,+\,
     \sum_{\bf q} (n_{a\bf q}+n_{b\bf q}+1)\,\tilde{\omega}_{\bf q}
\,.
\label{E_DOTTJ_EX}
\end{eqnarray}
The renormalized spin-wave energies $\tilde{\omega}_{\bf q}$ 
depend on the number of holes
$M=\delta\cdot N$, their momenta and spins $\{{\bf k}\sigma\}$, and the 
external field $B_A$.
Explicitly they are given by\cite{VojBeck1,VojBeck2}
\begin{eqnarray}
\tilde{\omega}_{\bf q}
   \,&=&\, {Jz_0 \over 2}(1-\delta)
        \,\left(1+\epsilon\,+\, \gamma({\bf q}) \nu_{\bf q} (1-\delta)\,\right)
        \,+\,{1\over 2}\lambda_1\,t\,F_1 (1-\delta)
        \,+\,\lambda_2\,{J\over 2}\,F_2 (1-\delta)^2 \nonumber\\
   \,&=&\, 2J\,(1-\delta) 
        \sqrt{ \Lambda^2
        \,-\,\gamma({\bf q})^2
        \,-\, \lambda_1\, {t\over J} \left( \Lambda F_1
        \,-\, \gamma({\bf q}) F_3 \right)
        }
\label{OMEGARES3}
\end{eqnarray}
with the substitutions $\tanh\mu_{\bf q}=\nu_{\bf q}$,
\begin{equation}
\Lambda \,=\, 1+\epsilon \,+\, {B_A \over 2J(1-\delta)}
\end{equation}
and
\begin{eqnarray}
t\, F_1\, &=& \,\sum_{m=1}^{M} \,\,
  \langle m|(a_{{\bf q}} b_{-{\bf q}})^{\cdot}
   \, H_{t} A_{1}\,
  (a_{{\bf q}}^+ b_{-{\bf q}}^+)^{\cdot}
  |m\rangle^c \nonumber\\
{J\over 2}\,F_2 \,&=&\, \sum_{m=1}^{M}\, \,
  \langle m|(a_{{\bf q}} b_{-{\bf q}})^{\cdot}
  \, H_{\bot}\, A_{2}\,
  (a_{{\bf q}}^+ b_{-{\bf q}}^+)^{\cdot}
  |m\rangle^c \, \nonumber\\
t\, F_3\, &=& \,\sum_{m=1}^{M} \,\,
  \langle m|(a_{{\bf q}} b_{-{\bf q}})^{\cdot}
   \, H_{t} A_{1}\,
  |m\rangle^c
\,.
\label{FTERMS}
\end{eqnarray}
In the second step of (\ref{OMEGARES3}) we have inserted the solution 
for $\nu_{\bf q}$ obtained from eq. (\ref{DOP_QUADGL}).
Due to the approximation of independent holes the expectation values 
with $|\phi_0\rangle$ in (\ref{E_DOTTJ_EX},\ref{OMEGARES3}) factorize 
into contributions from each of the holes.
So the cumulant expectation values in (\ref{FTERMS}) have to be taken with 
one-hole states $|m\rangle$ defined by
\begin{equation}
|m\rangle = c_{{\bf k}_m \sigma_m} |\phi_{N{\mathaccent 19 e}el}\rangle .
\end{equation}
The sum over $m$ runs over all holes in the system.
The expression (\ref{OMEGARES3}) includes all terms up to first order
in the hole concentration $\delta$, i.e., hole-hole interactions
are neglected.

To calculate the magnetization of the system according to (\ref{MAG_FDIFF})
we have to sum up the contributions from all states. 
The partition function is given by
\begin{eqnarray}
Z  \,\,=\,\, 
   \sum_{\{{\bf k}\sigma\}}
   \sum_{\{n_{a\bf q},n_{b\bf q}\}}
   e^{-\beta E_{\{{\bf k}\sigma\},\{n_{a\bf q},n_{b\bf q}\}} }
\,.
\end{eqnarray}
In the partition function we can exactly sum the contributions from the
excited spin waves which are bosons. Note that the spin-wave excitations
do not affect the hole part because of our rigid-band approximation.
Thus we can define a probability $p_{\{{\bf k}\sigma\}}$ for one-hole momentum
distribution $\{{\bf k}\sigma\}$:
\begin{eqnarray}
p_{\{{\bf k}\sigma\}} \,\,&=&\,\, {1 \over Z}
  \sum_{\{n_{a\bf q},n_{b\bf q}\}}
  e^ {-\beta\, E_{\{{\bf k}\sigma\},\{n_{a\bf q},n_{b\bf q}\}} } \nonumber\\
 \,\,&=&\,\,
    \exp ( -\beta\, \sum_m \epsilon_{{\bf k}_m} )
    \prod_{\bf q} (\sinh \beta\tilde{\omega}_{\bf q}/2)^{-2}
  \over
    \sum_{\{ {\bf k}\sigma\}' }
    \exp ( -\beta\, \sum_{m'} \epsilon_{{\bf k}_{m'}} )
    \prod_{\bf q} (\sinh \beta\tilde{\omega}_{\bf q}/2)^{-2}
\end{eqnarray}
where we have used
\begin{equation}
  \sum_{n=0}^{\infty} e^{-\beta (n\,+\,{1 \over 2})\tilde{\omega}}
\,\,=\,\,
  {1\over \sinh \beta\tilde{\omega}/2}
\,.
\end{equation}

Finally, one obtains the following expression for the staggered magnetization
(\ref{MAG_FDIFF}):
\begin{equation}
M_{eff} \,\,=\,\, {1 \over 2} \,\,-\,\, \delta M_{holes} \,\,-\,\,
  \sum_{\{{\bf k}\sigma\}} p_{\{{\bf k}\sigma\}} \, {1\over N}
  \sum_{\bf q} \left({\partial \tilde{\omega}_{\bf q} \over \partial B_A} \,
  \cosh {\beta\tilde{\omega}_{\bf q} / 2}
   \,\,-\,\,1 \right)
\label{MAGTJ_RES}
\end{equation}
where $\delta M_{holes}$ is the contribution of the hole quasiparticles
to the decrease of the magnetization. Within the rigid-band approximation
it is independent of the temperature. It is determined by the size
of the quasiparticles (spin bags) and can be expressed by the hole
concentration $\delta$ and the path
coefficients $\lambda_n$ of our ansatz (\ref{UTJ1}):
\begin{equation}
\delta M_{holes} \,\,=\,\, \delta \cdot
  {\sum_n n \lambda_n^2 \over \sum_n \lambda_n^2}
\,.
\end{equation}

The last term in (\ref{MAGTJ_RES}) contributing to the magnetization
is calculated numerically.
The integral over the spin-wave momenta has to be treated carefully for
small $\bf q$. 
For an isotropic system ($\epsilon=0$) and finite temperature it diverges
logarithmically indicating an instability of the magnetic order in
accordance with Mermin-Wagner theorem\cite{MerminWagner}.
The sum over the hole configurations $\{\bf k\}$ has been calculated 
for a finite lattice. 
The contributions from states with hole excitations, i.e., states with
a hole momentum distribution different from the ground state, are rather 
small for the low temperatures considered here.

In Fig. 1 we show the staggered magnetization calculated from 
(\ref{MAGTJ_RES}) for $t/J=5$ and
an anisotropy $\epsilon=2*10^{-6}$ in dependence on temperature and hole
concentration. At $T=0$ and $\delta=0$ it has a value of about 61\% of its
saturation value which is of course identical with the result obtained
in linear spin-wave theory. The small anisotropy does not have an
essential influence on this value. For increasing temperature and/or 
increasing doping level the magnetization decreases and reaches zero at a
line in the $T-\delta$-plane. This line shown in Fig. 3 can be interpreted 
as the boundary of the antiferromagnetic phase where we expect a 
second-order phase transition to a paramagnetic phase.
Note that linear spin-wave approximation becomes questionable for
vanishing sublattice magnetization. Therefore, the results presented here
are reasonable in the region of small doping and low temperature, but
the actual calculation can not provide a reliable description of
the system in the vicinity of the magnetic phase transition
($T \approx T_N$) and beyond it. Our starting point was the N\'{e}el state, 
and spin fluctuations are included by a perturbational method based on
cumulants, but we always describe a system
with antiferromagnetic long-range order\cite{VojBeck1}.
The behavior of the zero-temperature
magnetization is shown in Fig. 2 for $\epsilon=0$, for a discussion see
\cite{VojBeck2}.


\section{Conclusion}

The aim of this work was to present a method for calculating
thermodynamic properties in correlated electronic systems. This method
is based on the introduction of cumulants and is an extension of
a cumulant approach for ground-state calculations which has been 
applied in the past to a wide range of strongly correlated systems.

After a short review of the ground-state version we have developed
the extension of the cumulant method for the calculation of excitation
energies. It is based on the introduction of an unitary operator $U$
transforming the set of unperturbed eigenstates of $H_0$ into the full
eigenstates of $H$. For the unitary operator we use an appropriate 
exponential ansatz. Having calculated the spectrum of excitation energies
we obtain the free energy and other thermodynamical quantities. The
method is especially appropriate for low-temperature expansions if
one has to take into account only low-lying excitations of the 
considered system.

In the second part of this paper we have applied our method 
to the weakly doped $t$-$J$ model with an anisotropic exchange at 
finite temperature. Based on the investigation
of the renormalization of the spin-wave spectrum by mobile holes
\cite{VojBeck1,VojBeck2} we have calculated the staggered magnetization
as function of doping, temperature and magnetic anisotropy.
These calculations are only valid within the antiferromagnetic
phase of the $t$-$J$ model at small doping concentrations and low temperatures
because of the use of linear spin-wave approximation.
Possible improvements could include magnon-magnon
interaction, i.e. nonlinear spin-wave theory, which leads to a 
temperature-dependent spin-wave spectrum.

Summarizing, we have developed a general formalism for the investigation
of thermodynamic properties in both weakly and strongly correlated
many-body systems.

\acknowledgments
{It is a pleasure for us to thank W. Brenig and K. Kladko for 
helpful discussions.}


\appendix

\section{Generalized cumulants}

It was proposed by Kladko\cite{Kladko} to define cumulant expectation 
values with different bra and ket vectors. They are defined in analogy
to (\ref{CUM_DEF}) by
\begin{equation}
\langle\Phi|\prod_i A_i^{n_i}|\phi\rangle^c\, = \,
\left(\prod_i \left({\partial \over \partial\lambda_i} \right)^{n_i} \right)
\ln\,\langle\Phi|\prod_i {\rm e}^{\lambda_i A_i}|\phi\rangle\,
|_{\lambda_i=0\,\forall\,i}\,.
\label{GENCUM_DEF}
\end{equation}
The vectors $\langle\Phi|$ and $|\phi\rangle$ must have a non-zero overlap,
$\langle\Phi|\phi\rangle \neq 0$. It is easy to show that the cumulant equations
(\ref{11},\ref{13}) and (\ref{15},\ref{16}) are also valid with a
bra vector $\langle\Phi|$ different from $|\phi_n\rangle$, e.g.
\begin{eqnarray}
  E_n &=&  \langle\Phi| H U |\phi_n\rangle^c   \\
   0  &=&  \langle\Phi| (S_{\mu} - S_{\mu}^{\dagger})^{\dagger}
               H U |\phi_n\rangle^c
\end{eqnarray}
with $\langle\Phi|\phi_n\rangle \neq 0$. These equations follow from
the fact that $U$ transforms the unperturbed state $|\phi_n\rangle$ into
an eigenstate of the full Hamiltonian $H$.

\section{Evaluation of cumulant expectation values}

In this appendix we show how to evaluate cumulants containing an exponential
ansatz for the wave operator $\Omega$ or the unitary operator $U$.
The basic relation written with generalized cumulants is
\begin{equation}
\langle\Phi|\,\prod_i A_i^{n_i}\,e^S\,|\phi\rangle^c\, = \,
{\langle\Phi|\prod_i A_i^{n_i}|e^S \phi\rangle^c }
\label{KUMREL1}
\end{equation}
with $A_i$ and S being arbitrary operators and 
$\langle\Phi|\phi\rangle \neq 0$.
Note that on the l.h.s. the operators $S$ and $A_i$ are subject to
cumulant ordering whereas on the r.h.s. only the $A_i$ operators are
cumulant entities. However, the cumulants on the r.h.s. are formed
with the new ket vector $|e^S \phi\rangle$.

Eq. (\ref{KUMREL1}) can be proved either by integrating infinitesimal transformations
$(1+S/N)$ and using properties of cumulants\cite{Kladko}
or by explicit use of the definition of cumulant expectation values.
Here we demonstrate the second way. Starting from the definition of
generalized cumulant expectation values (\ref{GENCUM_DEF}) for
a product of arbitrary operators $A_i$ and arbitrary states 
$\langle\Phi|$, $|\phi\rangle$ with $\langle\Phi|\phi\rangle \neq 0$ 
we consider the following expression:
\begin{eqnarray}
&&\langle\Phi| \prod_i A_i^{n_i} \, e^{\alpha S} |\phi\rangle^c
  \nonumber \\
&&\, = \,
\sum_{n=0}^\infty {\alpha^n \over {n!}} 
\langle\Phi| \prod_i A_i^{n_i} S^n |\phi\rangle^c \nonumber\\
&&\, = \,
\sum_{n=0}^\infty {\alpha^n \over {n!}} \,\,
\left({\partial \over \partial\xi} \right)^n
\left[
\left(\prod_i \left({\partial \over \partial\lambda_i} \right)^{n_i} \right)
\ln\langle\Phi| \prod_i e^{\lambda_i A_i} e^{\xi S}|\phi\rangle\,
\right ]_{\xi=0 \atop \lambda_i=0\,\forall\,i}\,.
\end{eqnarray}
The last expression can be interpreted as a series expansion 
with respect to $\xi$ around 0 of the term in the brackets $[...]$:
\begin{eqnarray}
\langle\Phi| \prod_i A_i^{n_i} \, e^{\alpha S} |\phi\rangle^c
\, &=& \,
\left(\prod_i \left({\partial \over \partial\lambda_i} \right)^{n_i} \right)
\ln\langle\Phi| \prod_i e^{\lambda_i A_i} e^{\alpha S}|\phi\rangle\,
|_{\lambda_i=0\,\forall\,i} \nonumber \\
\, &=& \,
 \langle\Phi| \,\, \prod_i A_i^{n_i} \,\,
 | e^{\alpha S} \phi\rangle^c
\,.
\end{eqnarray}
In the last equation we have reintroduced generalized cumulants, now formed with
the bra state $\langle \Phi|$ and the ket state
$| e^{\alpha S} \phi\rangle$. With $\alpha=1$ we obtain
the desired result (\ref{KUMREL1}).

Explicitly we find from (\ref{KUMREL1}):
\begin{eqnarray}
\langle\Phi|\, A\, e^S\,|\phi\rangle^c\, &=& \,
  { \langle\Phi|\, A\, e^S\,|\phi\rangle \over
    \langle\Phi|\, e^S\,|\phi\rangle } \,, \nonumber\\
\langle\Phi|\, AB\, e^S\,|\phi\rangle^c\, &=& \,
  { \langle\Phi|\, AB\, e^S\,|\phi\rangle \over
    \langle\Phi|\, e^S\,|\phi\rangle }
\,-\,
  { \langle\Phi|\,A\, e^S\,|\phi\rangle \,
    \langle\Phi|\,B\, e^S\,|\phi\rangle \over
    \langle\Phi|\,    e^S\,|\phi\rangle^2 } \, ,
\end{eqnarray}
and so on.

\section{Comparison of the zero-temperature cumulant approach with other methods}

For practical calculations the zero-temperature cumulant method together
with the exponential ansatz for the wave operator $\Omega$ consists 
of selecting an appropriate set of
operators $S_\nu$, i.e., writing down an ansatz for the ground-state
wavefunction. Then the coefficients $\alpha_\nu$ are determined using the
equations (\ref{13}). The main advantage of this procedure compared to
other methods is that the exponential term occurs only once in all equations.

In a standard variational calculation one uses an ansatz for the wavefunction
and minimizes the ground-state energy by variation of the coefficients.
In such a calculation the ansatz wavefunction (including the exponential
operator) usually occurs four times,
$E_0 = \langle\psi_0|H|\psi_0\rangle \,/\, \langle\psi_0|\psi_0\rangle$.
Furthermore, a wavefunction with an exponential ansatz usually cannot be
normalized. So both numerator and denominator of the energy expression 
might diverge with an exponential
of the system size whereas their ratio should be proportional to the system 
size.

The physical difference between both methods is the following: 
In a variational calculation the aim is minimizing the total energy
of the system whereas in the cumulant method the aim is finding an
eigenstate of $H$. (Note that eq. (\ref{13}) is exactly the
condition of $e^S |\phi_0\rangle$ being an eigenstate of $H$.)

There is a close relationship of the equations (\ref{11},\ref{12},\ref{13}) to the
so-called coupled cluster method.
This approach which was originally invented for studies in nuclear physics
is also size consistent and does not involve
Wick's theorem. For a review see Bishop\cite{Bishop}. Recently it was 
shown\cite{SchorkFul92} that the coupled-cluster method can be derived from 
the cumulant expressions (\ref{11}) and (\ref{13}). 
Comparing practical calculations the cumulant method with an 
exponential ansatz is again easier to handle 
than the coupled-cluster scheme because the exponential term occurs 
only once in the cumulant equations and twice in the coupled-cluster
equations.

Usually these different methods lead to different (approximate) results
when calculating ground-state quantities. However, if the ansatz for the
ground-state wavefunction covers the exact ground state, i.e., if the 
subspace spanned by the operators $S_\nu$ contains the exact ground-state
wavefunction, then of course all methods lead to the same exact result.
In the following we briefly show how to derive coupled-cluster and variational
equations from the cumulant method if one assumes that the
exact ground state has the form $|\psi_0\rangle = e^S |\phi_0\rangle$ with
$S=\sum_\nu \alpha_\nu S_\nu$.
We note that eq. (\ref{13}) also holds for arbitrary composite operators, e.g.,
$0=\langle\phi_0|A B H \Omega |\phi_0\rangle^c$ for arbitrary operators
$A$ and $B$.
Inserting (\ref{13}) into (\ref{11}) one obtains
\begin{eqnarray}
E_0 \:=\: \langle\phi_0|H e^S | \phi_0\rangle^c
    \:=\: \langle\phi_0|e^{-S} H e^S | \phi_0\rangle^c
    \:=\: \langle\phi_0|{e^S}^\dagger H e^S | \phi_0\rangle^c
\,.    
\end{eqnarray}
Evaluating the cumulants in analogy to appendix B leads to    
\begin{eqnarray}
E_0
  \:= \:\langle\phi_0|e^{-S} H e^S | \phi_0\rangle
  \:=\: {
   {\langle\psi_0| H | \psi_0\rangle}
   \over 
   {\langle\psi_0|\psi_0\rangle}
   }
\,.    
\end{eqnarray}
These are the energy expressions for the coupled-cluster and the variational
scheme, respectively.
The equations for the coefficients are obtained from (\ref{13}) as
follows:
\begin{eqnarray}
0 \:=\: \langle\phi_0|S_\nu^\dagger H e^S | \phi_0\rangle^c
  \:=\: \langle\phi_0|S_\nu^\dagger e^{-S} H e^S | \phi_0\rangle^c 
  \:=\: \langle\phi_0|{e^S}^\dagger S_\nu^\dagger H e^S | \phi_0\rangle^c
\,.
\end{eqnarray}
Transforming again the cumulants and using $\langle\phi_0|S_\nu|\phi_0\rangle=0$
one finds    
\begin{eqnarray}
0  \:= \:\langle\phi_0| S_\nu^\dagger e^{-S} H e^S | \phi_0\rangle
\end{eqnarray}
and
\begin{eqnarray}
0 \:=\:
   \langle\psi_0| S_\nu^\dagger H | \psi_0\rangle^c
   \:+\:
   \langle\psi_0| H S_\nu | \psi_0\rangle^c
  \:=\:
  \frac {\partial} {\partial\alpha_\nu} \,E_0
\,.
\label{VAR_COEFFEQ}
\end{eqnarray}
These two conditions are the equations for the coefficients $\alpha_\nu$
within the coupled-cluster and the variational method. The second step of
(\ref{VAR_COEFFEQ}) includes evaluating the new cumulants with $|\psi_0\rangle$
yielding exactly the four terms arising from the differentiation of the 
energy expression 
$\langle\psi_0| H | \psi_0\rangle \,/\, \langle\psi_0|\psi_0\rangle$ 
with respect to $\alpha_\nu$.

We want to note here that the wave operator (\ref{10}) of the cumulant 
approach is not limited to an exponential form (as is the case e.g. in 
coupled-cluster calculations). So the cumulant method appears to be
the more general and powerful scheme for the calculation of
ground-state properties.
For modified applications of the cumulant approach see e.g.\cite{Polatsek}.

\section{Unitary transformation for the Heisenberg antiferromagnet}

In this appendix we consider the Bogoliubov transformation which
diagonalizes the Heisenberg antiferromagnet within linear spin-wave 
theory. We want to prove the equivalence of the forms of $U_{\bf q}$
given in (\ref{UHEIS1}) and (\ref{UHEIS2}). We have to prove
\begin{eqnarray}
&&\exp(\mu(a^+ b^+\,-ab))\ \nonumber \\
&&\,= \quad
\exp ((\tanh \mu)a^+ b^+)\,
\exp\left (-(\ln\, \cosh \mu)(a^+a+b^+b+1) \right ) \,
\exp(-(\tanh \mu)a b) \label{BOGOLID2}
\end{eqnarray}
where we used the short-hand notations $a$ and $b$ for the boson operators
$a_{\bf q}$ and $b_{\bf q}$.

Note that the left-hand side of (\ref{BOGOLID2}) is the unitary transformation
$U_{\bf q}$ (for fixed momentum $\bf q$) transforming the "original" bosons
$a$ and $b$ into the "new" bosons $\alpha$ and $\beta$,
\begin{eqnarray}
U_{\bf q}\,a\,U_{\bf q}^\dagger \quad=&\quad \alpha \quad=\quad
  a\,\cosh \mu\,-\,b^+\,\sinh\mu \, , \nonumber\\
U_{\bf q}\,b\,U_{\bf q}^\dagger \quad=&\quad \beta \quad=\quad
  b\,\cosh \mu\,-\,a^+\,\sinh\mu \, .
\end{eqnarray}
These relations can be easily checked by expanding both sides with respect to $\mu$.
With this we obtain
\begin{eqnarray}
U_{\bf q}\,ab\quad=\quad\cosh^2\mu
   \left ( ab\,-\,\tanh\mu\,(a^+a+b^+b+1) \,+\,(\tanh^2\mu)\,a^+b^+
   \right )\,U_{\bf q}
\,. \label{BOGOLID3}
\end{eqnarray}
Another commutation relation needed is
\begin{eqnarray}
\exp ((\tanh \mu)a^+ b^+)\,\,a^+a \quad=\quad
\left (a^+a\,-\,(\tanh\mu)a^+b^+ \right )\,\exp ((\tanh \mu)a^+ b^+)
\, . \label{BOGOLID4}
\end{eqnarray}
which can be derived from a well-known identity of Kubo\cite{Kubo59} for
arbitrary operators $A$ and $H$:
\begin{eqnarray}
[A,e^{-\lambda H}]\quad=\quad
- \int_0^\lambda d u \,\,e^{-(\lambda - u)H}\, [A,H]\, e^{-uH}
\end{eqnarray}
Now we can prove (\ref{BOGOLID2}). 
We first multiply (\ref{BOGOLID2}) by $\exp(-(\tanh \mu)a b)$ from
the right in order to define the two functions
\begin{eqnarray}
f_l(\mu) \quad&=&\quad \exp(\mu(a^+ b^+\,-ab))\,\,
         \exp(-(\tanh \mu)a b) \nonumber\\
f_r(\mu) \quad&=&\quad \exp((\tanh \mu)a^+ b^+)\,\,
         \exp \left (-(\ln\, \cosh \mu)(a^+a+b^+b+1) \right )
\, .
\end{eqnarray}
The next step is to compare the equations of motion of $f_l$ and $f_r$
with respect to $\mu$. One immediately finds
\begin{eqnarray}
{\partial \over \partial \mu} f_l\quad&=&\quad
   (a^+b^+\,-\,ab)\,f_l \,\,+\,\, f_l\, {ab \over \cosh^2 \mu} \,, \nonumber\\
{\partial \over \partial \mu} f_r\quad&=&\quad
    {a^+b^+ \over \cosh^2 \mu}\,f_r \,\,-\,\, 
    f_r\,(a^+a \,+\, b^+b \,+\, 1)\,\tanh\mu
\,. \label{BOGOLID5}
\end{eqnarray}
Now we insert (\ref{BOGOLID3}) and (\ref{BOGOLID4}) into
(\ref{BOGOLID5}) leading to:
\begin{eqnarray}
{\partial \over \partial \mu} f_l\quad&=&\quad
   \left ( a^+b^+\,-\,(\tanh\mu)\,(a^+a+b^+b+1) \,+\,(\tanh^2\mu)\,a^+b^+
   \right )\, f_l\, ,  \nonumber\\
{\partial \over \partial \mu} f_r\quad&=&\quad
    \left ( {a^+b^+ \over \cosh^2 \mu}
    -\,(\tanh\mu)\,(a^+a+b^+b+1) \,+\,(\tanh^2\mu)\,2a^+b^+
    \right )\,f_r
\,.
\end{eqnarray}
Using $\cosh^{-2}\mu\,+\,\tanh^2\mu\,=\,1$ we see that both expressions
are indeed identical. Having shown that $f_l(\mu)$ and $f_r(\mu)$ obey
the same equation of motion and the same initial condition
$f_l(0)=f_r(0)=1$ we have proved Eq. (\ref{BOGOLID2}).

\section{Calculation of expectation values for the Heisenberg antiferromagnet}

Here we demonstrate the evaluation of the energy expression (\ref{E_N_HEIS}).
For the denominator we find by use of (\ref{UHEIS2}):
\begin{eqnarray}
\langle\phi_{\{n_{a\bf q},n_{b\bf q}\}}| U
       |\phi_{\{n_{a\bf q},n_{b\bf q}\}}\rangle \,\,=\,\,
\prod_{\bf q} \sum_{i=0}^{min(n_{a\bf q},n_{b\bf q})}
   \sqrt {1-\nu_{\bf q}^2} ^ {\,n_{a\bf q}+n_{b\bf q}+1-2i} \,
   (-\nu_{\bf q}^2)^i \,
   \left( \stackrel {n_{a\bf q}}{i} \right) \,
   \left( \stackrel {n_{b\bf q}}{i} \right)
\,.
\end{eqnarray}
For the numerator of (\ref{E_N_HEIS}) one obtains
\begin{eqnarray}
\langle\phi_{\{n_{a\bf q},n_{b\bf q}\}}| H_0\,U
      |\phi_{\{n_{a\bf q},n_{b\bf q}\}}\rangle \,\,= \,\,
\left(E_{N{\mathaccent 19 e}el} \,+\,
      2J \sum_{\bf k} (n_{a\bf k}+n_{b\bf k})\,
\right)\,
\langle\phi_{\{n_{a\bf q},n_{b\bf q}\}}| U
      |\phi_{\{n_{a\bf q},n_{b\bf q}\}}\rangle
\end{eqnarray}
and
\begin{eqnarray}
\langle\phi_{\{n_{a\bf q},n_{b\bf q}\}}| H_1\,U
      |\phi_{\{n_{a\bf q},n_{b\bf q}\}}\rangle \,\,=
      \quad\quad\quad\quad\quad\quad\quad\quad\quad\quad
      \quad\quad\quad\quad\quad\quad\quad\quad\quad \nonumber\\
\sum_{\bf k} \frac {J z_0} {2} \gamma({\bf k}) \,
\left (
\sum_{i=0}^{min(n_{a\bf k},n_{b\bf k})}
   \sqrt {1-\nu_{\bf k}^2} ^ {\,n_{a\bf k}+n_{b\bf k}+1-2i} \,
   (-\nu_{\bf k})^i \,\times
   \right . \quad\quad\quad\quad\quad\quad \nonumber\\
\left. 
   \left( \nu_{\bf k}^{i-1}i \,+\, \nu_{\bf k}^{i+1}
          \frac {(n_{a\bf k}+1)(n_{b\bf k}+1)} {i+1}\right )
   \left( \stackrel {n_{a\bf k}}{i} \right) \,
   \left( \stackrel {n_{b\bf k}}{i} \right)
\right ) \,\times \nonumber\\
\prod_{{\bf q} \neq {\bf k}} \sum_{i=0}^{min(n_{a\bf q},n_{b\bf q})}
   \sqrt {1-\nu_{\bf q}^2} ^ {\,n_{a\bf q}+n_{b\bf q}+1-2i} \,
   (-\nu_{\bf q}^2)^i \,
   \left( \stackrel {n_{a\bf q}}{i} \right) \,
   \left( \stackrel {n_{b\bf q}}{i} \right)
\quad\quad\quad
\, .
\label{HEISAPP3}
\end{eqnarray}
After straightforward algebra we find from (\ref{HEISAPP3}):
\begin{eqnarray}
\langle\phi_{\{n_{a\bf q},n_{b\bf q}\}}| H_1\,U
      |\phi_{\{n_{a\bf q},n_{b\bf q}\}}\rangle \,\,= \,\,
\sum_{\bf k} \frac {J z_0} 2 \gamma({\bf k}) \nu_{\bf k} \,
(n_{a\bf k}+n_{b\bf k}+1)\,
\langle\phi_{\{n_{a\bf q},n_{b\bf q}\}}| U
      |\phi_{\{n_{a\bf q},n_{b\bf q}\}}\rangle
\, .
\end{eqnarray}
Collecting all terms leads directly to (\ref{EHEIS2}).

\begin{figure}
\epsfxsize=10.6cm
\epsfysize=8cm
\epsffile{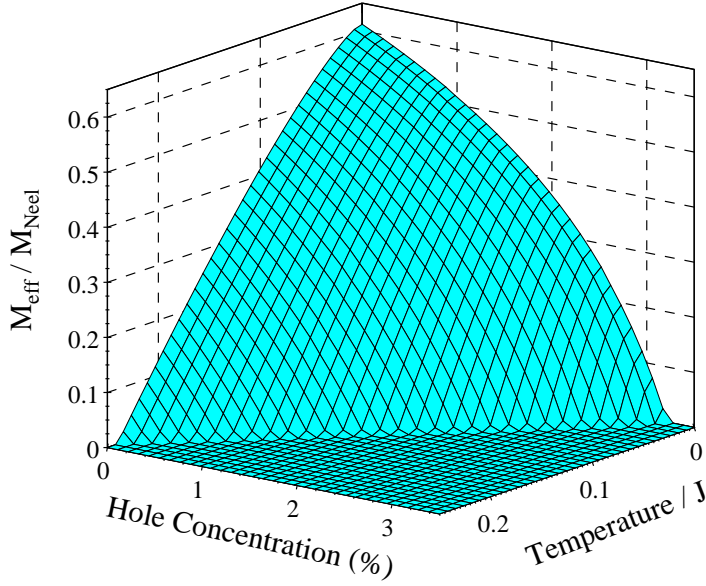}
\vspace{0.5cm}
\caption{Staggered magnetization as function of hole concentration $\delta$
and temperature $T$ for different values of the anisotropy $\epsilon$.
$t/J=5, \epsilon = 2*10^{-6}$.}
\end{figure}

\begin{figure}
\epsfxsize=10.6cm
\epsfysize=8cm
\epsffile{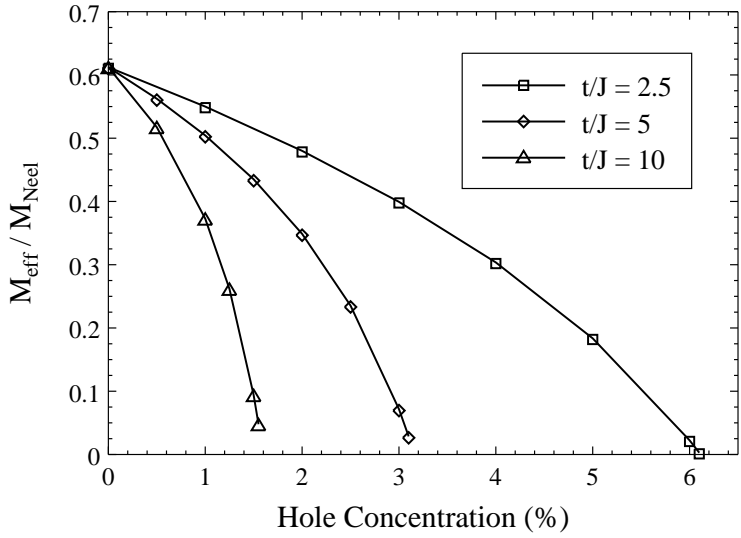}
\vspace{0.5cm}
\caption{Staggered magnetization as function of hole concentration $\delta$
for different values of $t/J$ at zero temperature. The magnetic anisotropy
$\epsilon$ is set to zero.}
\end{figure}

\begin{figure}
\epsfxsize=10.6cm
\epsfysize=8cm
\epsffile{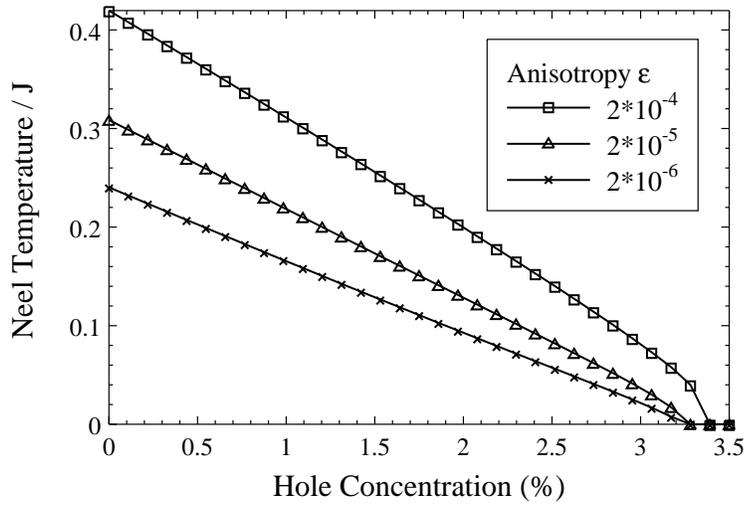}
\vspace{0.5cm}
\caption{Boundary of antiferromagnetic phase for $t/J=5$ and different
values of the magnetic anisotropy $\epsilon$. This boundary is given by the
condition $M_{eff}=0$.}
\end{figure}

\end{document}